  \documentclass[twocolumn,prl,aps]{revtex4}

 \usepackage[dvips]{graphicx}
 \usepackage[dvips]{graphics}

\newlength{\bxwidth}\bxwidth=0.8\textwidth

\begin{document}
\title{Two resonant magnetic modes in an overdoped high-$\bf T_c$ superconductor}
\author{S. ~Pailh\`es$^1$, Y. ~Sidis$^1$, P.~Bourges$^{1\ast}$, C. Ulrich$^2$, V.
Hinkov$^2$, L.P.~Regnault$^3$, A. ~Ivanov$^4$, B. Liang$^2$, C.T.
Lin$^2$, C. Bernhard$^2$, and B. Keimer$^{2}$.}

\affiliation{$^1$ Laboratoire L\'eon Brillouin, CEA-CNRS, CE-Saclay, 91191 Gif sur Yvette, France.\\
$^2$ Max-Planck-Institut f\"ur Festk\"orperforschung, 70569 Stuttgart, Germany\\
$^3$ CEA Grenoble, DRFMC, 38054 Grenoble cedex 9, France.\\
$^4$ Institut Laue Langevin, 156X, 38042 Grenoble cedex 9, France.
}


\begin{abstract}

A detailed inelastic neutron scattering study of the overdoped
high temperature copper oxide superconductor
${Y_{0.9}Ca_{0.1}Ba_{2}Cu_3O_{7}}$ reveals two distinct magnetic
resonant modes in the superconducting state. The modes differ in
their symmetry with respect to exchange between adjacent copper
oxide layers. Counterparts of the mode with odd symmetry, but not
the one with even symmetry, had been observed before at lower
doping levels. The observation of the even mode resolves a
long-standing puzzle, and the spectral weight ratio of both modes
yields an estimate of the onset of particle-hole spin-flip
excitations.

\end{abstract}

\maketitle

In  various high-$\rm T_c$ superconductors with crystal structures
comprised of $\rm CuO_2$ monolayer units (${\rm
Tl_2Ba_2CuO_{6+\delta}}$\cite{tl2201}) and bilayer units ($\rm
YBa_{2}Cu_3O_{7}$ (YBCO)
\cite{rossat,mook93,fong95,hoechst,fong00,dai01} and ${\rm
Bi_2Sr_2CaCu_2O_{8+\delta}}$ (BSCO) \cite{fong99}), the magnetic
excitation spectrum in the superconducting (SC) state is dominated
by a resonant mode. For a given compound, the excitation vanishes
at $\rm T_c$ without any marked change of the excitation energy
${\rm E_r}$. A comparison of different compounds has uncovered the
scaling relation $ E_r \sim 5 k_B T_c$ \cite{he01}. Signatures of
a strong interaction of the mode with charged quasiparticles
\cite{abanov02}  have been observed in various spectroscopic data
including angle-resolved photoemission spectroscopy
\cite{eschrig03} and tunneling \cite{zasadzinski01}.

The resonant mode occurs around the planar wave vector
$Q_{AF}=(\pi/ a, \pi / a)$ characteristic of antiferromagnetic
(AF) fluctuations in both monolayer and bilayer systems. ($a$
stands for the Cu-Cu distance in the ${\rm CuO_2}$ planes). Due to
interlayer exchange coupling, bilayer systems are additionally
expected to exhibit non-degenerate magnetic excitations with odd
($o$) and even ($e$) symmetry with respect to exchange of the
layers. While this has been confirmed in insulating YBCO
\cite{reznik96} by the observation of acoustic and optical spin
waves (whose symmetry is $o$ and $e$, respectively), one of the
most puzzling features of the resonant mode in superconducting
bilayer systems is that it appears exclusively in the odd channel.
On general grounds, both odd and even parts of the spin
susceptibility are required as input for theories of a
superconducting pairing mechanism based on magnetism. Differences
in the spin dynamics in $o$ and $e$ channels can even lead to
inter- and intra-layer pairing states of different symmetry
\cite{Li01}. Despite various attempts in underdoped and optimally
doped samples, the even resonant mode has thus far not been
observed, presumably due to a much weaker intensity. Here we
report, for the first time, the observation of odd and even modes
in a superconducting YBCO sample in which the $\rm CuO_2$ planes
are overdoped through partial substitution of Ca$^{2+}$ for
Y$^{3+}$.

\begin{figure}[t]
\begin{tabular}{ p{4cm} p{4cm} }
\includegraphics[width=4cm,height=6cm]{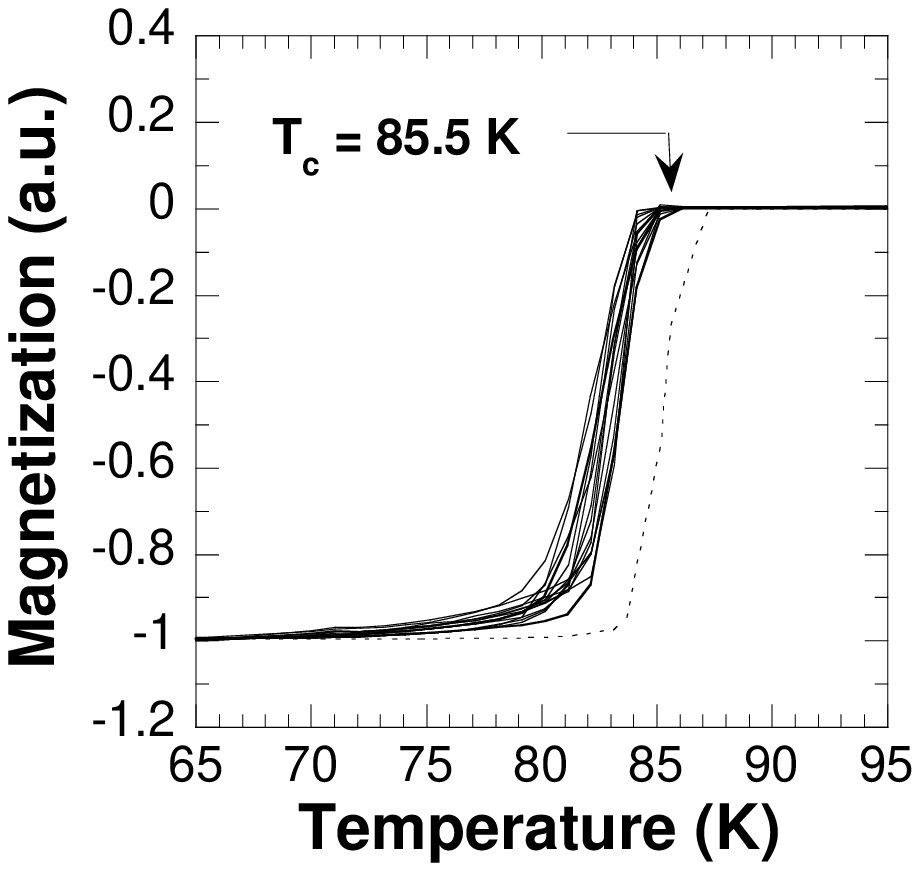} & \vspace{-6cm}\hspace{-.2cm}
\includegraphics[width=4cm,height=5cm]{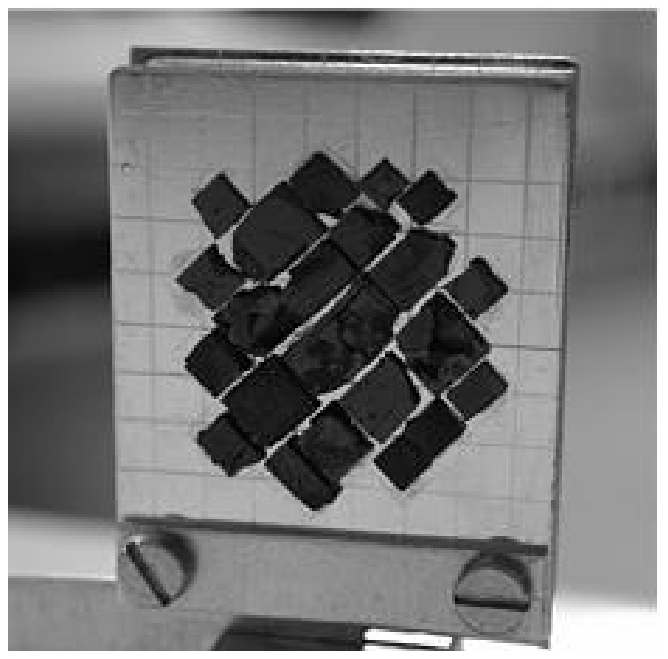}
\end{tabular} \caption {
{\label{fig1}} (left) Susceptibility of 15 individual ${\rm
Y_{0.9}Ca_{0.1}Ba_{2}Cu_3O_{7}}$ crystals (full curves).
For comparison, the dashed line represents the susceptibility
curve taken for one crystal after heat treatment to reduce the
oxygen content. Its higher $\rm T_c$ demonstrates that the samples
used in the neutron experiments are slightly overdoped. (right)
Photograph of the array of co-oriented overdoped single crystals.
60 crystals are glued onto Al plates only one of which is shown
for clarity.}
\end{figure}

Y$_{0.9}$Ca$_{0.1}$Ba$_{2}$Cu$_{{\bf 3}}$O$_{6+x}$ single crystals
were prepared by a top-seeded solution growth method described
previously \cite{YBCOCa_Lin}. About 60 single crystals with a
total volume of 350 mm$^{3}$ were co-aligned with a total
mosaicity of 1.4$^\circ$ as shown in Fig. \ref{fig1}a. Prior to
alignment, the Ca content and superconducting transition
temperature were determined by energy dispersive x-ray analysis
(EDX) and magnetometry, respectively, for each crystal
individually. The EDX measurements revealed an excellent
homogeneity of the Ca content both parallel and perpendicular to
the $\rm CuO_2$ layers, and the superconducting transitions
measured by magnetometry were sharp (full width $<$ 4K, Fig.
\ref{fig1}b). To achieve the overdoped state, the samples were
annealed in flowing oxygen at 500 $^{\circ}$C for 150h yielding $x
\simeq 1$. The total crystal mosaic has an onset $\rm T_c$ of 85.5
$\pm$ 0.6 K and calcium content of 10 $\pm$ 1 \% (mean $\pm$
standard deviation).

The measurements were taken at the 2T triple axis spectrometer at
the Laboratoire L\'eon Brillouin (Saclay, France), and at the IN8
triple axis spectrometer at the Institut Laue Langevin (Grenoble,
France). On 2T, a focusing pyrolytic graphite (PG) (002)
monochromator and analyzer were used, with a PG filter inserted
into the beam in order to eliminate higher order contamination.
The IN8 beam optics includes a vertically and horizontally
focusing Si (111) crystal as monochromator, and a PG (002)
analyzer. No filter was required on IN8 because the Si (222) Bragg
reflection is forbidden. The crystal was oriented such that
momentum transfers $Q$ of the form $Q=(H,H,L)$ were accessible. We
use a notation in which $Q$ is indexed in units of the tetragonal
reciprocal lattice vectors $2\pi/a=1.64 $\AA$^{-1}$ and
$2\pi/c=0.54 $\AA$^{-1}$.

As discussed previously \cite{fong00,liu95,millis96}, the cross
section for magnetic neutron scattering from bilayer cuprates
reads

\begin{eqnarray}
\frac{ \partial^2 \sigma (Q,\omega)}{\partial \Omega \partial
\omega} \propto F^2(Q) \Big[ \sin^2 (\pi z L)
 Im \lbrack \chi_{o}(Q,\omega)\rbrack + \nonumber\\
 \cos^2(\pi z L) Im \lbrack
\chi_{e}(Q,\omega)\rbrack \Big] \label{eq-bilayer}
\end{eqnarray}

\begin{figure}[t]
\centerline{\includegraphics[width=7 cm]{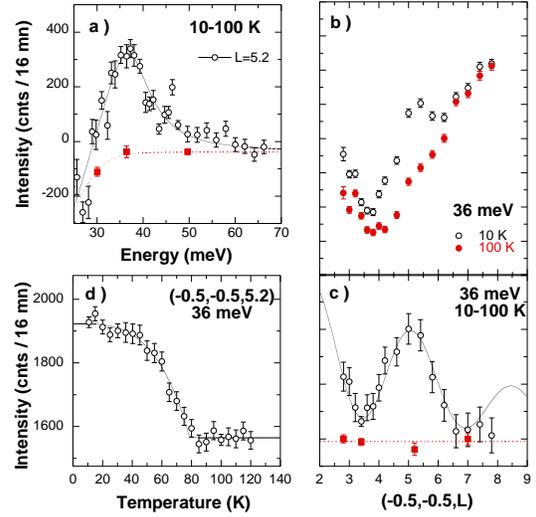}}
\caption{
{\label{fig2}} a) Difference of the neutron intensities measured
at T= 10 K ($<T_{c}$) and T= 100 K ($>T_{c}$) in the odd channel
at $Q=(0.5,0.5,5.2)$ versus excitation energy $E$.
b) Constant-energy scans at $E_r^o=36$ meV along the $L$-direction
perpendicular to the CuO$_2$ planes at 10 K and 100 K, c)
Difference  between the two scans shown in b). The full symbols in
a) and c) are determined by the difference of constant-energy
scans along $(H,H,L_0)$ direction for different $L_0$-values.
Dotted lines connecting these full symbols then correspond to the
reference level of magnetic scattering. d) Temperature dependence
of the neutron intensity
at the resonance energy $E=E_{r}^o$ and $Q=Q_{AF}$.}
\end{figure}

\noindent where ${\rm Im \lbrack \chi_{o,e}(Q,\omega) \rbrack}$ is
the imaginary part of the dynamical magnetic susceptibility in the
odd and even channels, respectively, {$z c=3.3$ \AA} is the
distance between ${\rm CuO_2}$ planes within a bilayer unit, and
$F(Q)$ is the magnetic form factor of Cu$^{2+}$. In order to
determine the energy of the magnetic resonant mode, we first
followed previous work \cite{fong00} and performed constant-$Q$
scans at $Q=(0.5,0.5,5.2)$ where the structure factor for odd
excitations is maximum. As previously found in YBCO$_7$
\cite{fong95,hoechst}, antiferromagnetic spin correlations are not
observable above the background level in the normal state, so that
the normal-state intensity can serve as a reference level. Fig.
\ref{fig2}a shows that the intensity increases in the
superconducting state, and that the difference between scans taken
in the superconducting and normal states exhibits the same
prominent peak that heralds the resonant mode in other copper
oxides. (Note that the reference level is negative and increases
with increasing energy for a given temperature difference, because
the background is predominantly determined by multiphonon
scattering events.) The peak energy, 36 meV, is lower than the
mode energy in optimally doped YBCO, but consistent with
observations in slightly overdoped BSCO \cite{he01} and with the
scaling relation $E_r \sim 5 k_{B}T_{c}$. Constant-energy scans at
36 meV along $(0.5,0,5,L)$ (Figs. \ref{fig2}b and \ref{fig2}c)
exhibit a sinusoidal intensity modulation that confirms the
previously observed odd symmetry of the resonant mode (Eq.
\ref{eq-bilayer}). In-plane constant-energy scans along
$(H,H,5.2)$ (not shown) exhibit a peak with an intrinsic full
width at half maximum of $\Delta Q = 0.36 \pm 0.05$ \AA$^{-1}$,
somewhat larger than $\Delta Q = 0.25\pm0.05$ \AA$^{-1}$
determined for the resonant mode in YBCO$_7$. The temperature
dependence of the peak intensity (Fig. \ref{fig2}d) shows that the
mode vanishes in the normal state, similarly to the optimally
doped systems.

\begin{figure}[t]
\centerline{\includegraphics[width=7 cm]{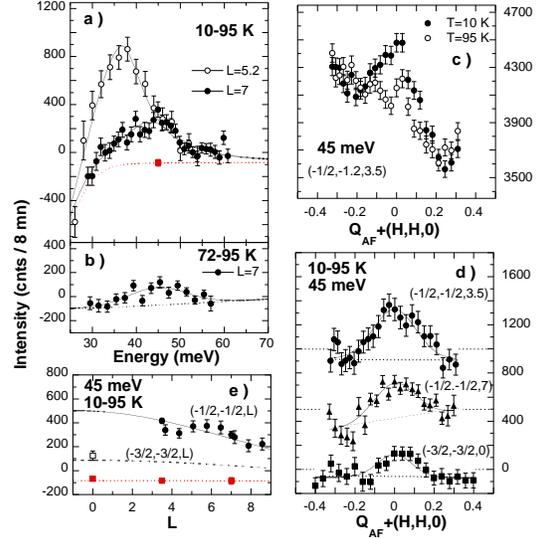}}
\caption{
{\label{fig3}} a, b) Constant-$Q$ scans, c, d)
constant-energy-scans
evidencing a second magnetic mode in the even channel around
$E_r^e \sim$ 43 meV. Panel c) shows raw neutron scans at two
different temperatures whereas a), b) and d) display differences
of the neutron intensities at T= 10 K ($<T_{c}$) and T= 100 K
($>T_{c}$). The two upper scans in d) were shifted by 300 counts
from the lower one for clarity, and the dotted line represents the
background level. e) Difference between constant energy-scans at
E=45 meV obtained at T= 10 K and T= 95 K along the $L$-direction
(open circles). The open square shows the magnetic intensity
measured by constant-energy scans at E=45 meV along $Q=(H,H,0)$
for H around 3/2. The full symbols, connected by the lower dotted
line, correspond to the measured reference level of the magnetic
scattering determined by constant energy-scans at E=45 meV along
$Q=(H,H,L_0)$ for different $L_0$-values. }
\end{figure}

While these observations are largely consistent with previous work
\cite{rossat,mook93,fong95,hoechst,fong00,dai01}, some new aspects
of the data of Fig. \ref{fig2}a are noteworthy. First, in contrast
to the resolution-limited energy profile in pure YBCO, the
resonant peak in the overdoped material exhibits a broader,
asymmetric line shape with a tail above $\rm E_r$ (Fig.
\ref{fig2}a). Second, the sinusoidal intensity modulation along
$L$ is not complete as some intensity remains at $L=3.6$ and 7.2
(Fig. \ref{fig2}c). Similar $L$-scans for YBCO$_{6.97}$ show the
full modulation \cite{hoechst}. In order to check for a possible
admixture of a second mode with even symmetry, we performed
constant-$Q$ scans at $Q=(0.5,0.5,7)$ where the structure factor
for even excitations (Eq. \ref{eq-bilayer}) is near its maximum.
The difference of the intensities below and above $\rm T_c$
(closed symbols in Fig. \ref{fig3}a) reveals a new mode around an
energy of 43 meV, clearly different from the mode energy in the
odd channel (open symbols in Fig. \ref{fig3}a). Such an even
excitation has not been observed in previous INS studies of YBCO
and BSCO, most likely because its intensity is smaller at lower
doping levels and hence indistinguishable from the background.
Further, the overdoped regime could be particularly suited to
detect such an even mode because electronic transport between
closely spaced $\rm CuO_2$ layers becomes coherent, as
demonstrated by recent experiments showing well-defined bonding
and antibonding bands \cite{bbs}.

\begin{figure}
\centerline{\includegraphics[width=8 cm,height=5
cm]{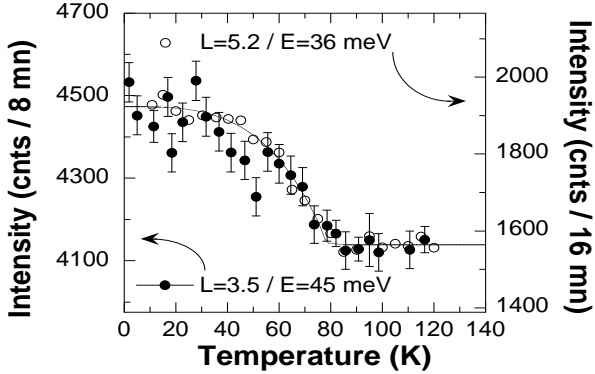}} \caption{
{\label{fig4}} Temperature dependence of the neutron intensity at
$Q_{AF}=(0.5,0.5,L)$ at the even mode energy $E_r^e=45$ meV and
$L=3.5$ (full circles and left scale)
The temperature dependence of the odd mode (open circles and right
scale) of Fig. \ref{fig2}.d, has been superposed without the error
bars for clarity.}
\end{figure}

In-plane constant-energy scans performed at maxima of the even
structure factor in different Brillouin zones (Figs. \ref{fig3}c
and \ref{fig3}d) show that the intensity is peaked at $Q_{AF}$.
The intensity decreases with increasing total wave vector $Q$,
following the anisotropic magnetic form factor of Cu$^{2+}$
\cite{shamoto} (Fig. \ref{fig3}e). This excludes phonons (whose
intensity generally increases with increasing $Q$) and
demonstrates the magnetic origin of the second mode. However, the
observed $L$-dependence at E=45 meV (Fig. \ref{fig3}e) differs
from the $\cos^2$ dependence expected for the even mode (Eq.
\ref{eq-bilayer}). As shown by Fig. \ref{fig3}a, the deviation
from the expected $\cos^2$ dependence is probably related to a
combined effect of the anomalous lineshapes of both modes and
their relative spectral weights.

The temperature dependences of the odd and even modes are closely
similar (Fig. \ref{fig4}). Moreover, Fig. \ref{fig3}b demonstrates
that the even mode remains peaked at 43 meV up to 72 K (= $T_c$ -
13 K). The peak amplitude of the even mode thus vanishes at $\rm
T_c$ without renormalization of the mode energy, as it does for
the odd mode. The findings underline the common origin of both
modes. The in-plane $Q$-width of the even mode, measured along the
(110) direction, is 0.45$\pm$0.05 \AA$^{-1}$, somewhat larger than
that of the odd mode (Fig. \ref{fig3}d). The energy profiles of
both modes are identical to within the experimental error.
Calibration of the magnetic intensity to that of an optical phonon
\cite{fong00} and integrating over the asymmetric profiles in
energy yields spectral weights of $W_o( Q_{AF}) = \int d \omega Im
\chi(Q_{AF},\omega) = 0.8 \mu_{B}^{2}$ per formula unit (f.u.) and
$W_e (Q_{AF}) = 0.36 \mu_{B}^{2}$/f.u. for odd  and even modes,
respectively. Averaged over the entire Brillouin zone, this
corresponds to $\langle W_o (Q) \rangle = 0.042 \mu_{B}^{2}$/f.u.
and $\langle W_e (Q) \rangle = 0.036 \mu_{B}^{2}$/f.u. These
values should be compared with $ W_o(Q_{AF}) =  1.6
\mu_{B}^{2}$/f.u. and $\langle W_o (Q) \rangle = 0.043
\mu_{B}^{2}$/f.u. for the odd mode in YBCO$_7$ \cite{fong00}. The
total magnetic spectral weight observed by neutron scattering is
thus preserved in the overdoped regime, but it is spread out over
a wider range of energy and momentum.

Two distinct magnetic modes of odd and even symmetry are thus
observed by spin-flip neutron scattering in the superconducting
state of overdoped YBCO, as predicted by mean-field theories of
d-wave superconductors in the presence of antiferromagnetic
interactions \cite{liu95,millis96,onufrieva02}. The observation
that the difference between the mode energies, $E_r^e - E_r^o
\simeq 7$ meV, is of the order of $ J_{\perp} \sim 10$ meV, the
interlayer superexchange coupling determined experimentally in
insulating YBCO \cite{reznik96}, is also consistent with these
models. Under the assumption that both modes are bound states in
the superconducting energy gap, their spectral weights are
predicted to be approximately proportional to their binding
energies, $\omega_c-E_r^{o,e}$ \cite{millis96}, so that

\begin{equation}
\frac{W_o (Q_{AF})}{W_e (Q_{AF})} =
\frac{\omega_c-E_r^o}{\omega_c-E_r^e}
\end{equation}

This yields an estimate of $\omega_{c} = 49$ meV for the threshold
of the continuum of particle-hole spin-flip excitations at the
wave vector $Q_{AF}$. The large spectral weight difference thus
implies a small binding energy of the even resonant mode. In the
limit of small binding energy, one further expects that the normal
state spectral weight below $\omega_{c}$ is only partially
redistributed into the bound state below $\rm T_c$, with the
remainder accumulating above the particle-hole threshold. This
could provide an explanation for the asymmetric line shape of the
mode. A larger distance of the resonant mode to $\omega_{c}$ would
also explain why this asymmetry is not observed at optimum doping.

In any case, both the existence of two distinct odd and even
excitations and their relative spectral weights put new
constraints on the various models to account for the magnetic
dynamics in the superconducting cuprates. Notably, our experiments
provide an estimate of the locus of the particle-hole spin-flip
excitations in the high-$\rm T_c$ cuprates, a direct measure of
the bulk superconducting energy gap $\Delta_k$.


 \normalsize{$^\ast$To whom correspondence should be addressed;
 E-mail: bourges@bali.saclay.cea.fr }

\end{document}